\documentstyle[twoside,fleqn,espcrc2,epsf]{article}

\newcommand{\AmS}{{\protect\the\textfont2
  A\kern-.1667em\lower.5ex\hbox{M}\kern-.125emS}}
\newcommand  {\equ}   {\!=\!}

\title{Making the gravitational path integral more Lorentzian\\
{\it or} \\
Life beyond Liouville gravity}

\author{{\bf R. Loll}\address{
        Max-Planck-Institut f\"ur Gravitationsphysik,
        Am M\"uhlenberg 1, D-14476 Golm, Germany},
        J. Ambj\o rn\address{The Niels Bohr Institute,
        Blegdamsvej 17, DK-2100 Copenhagen \O , Denmark},
        K.N. Anagnostopoulos\address{Department of Physics, 
        University of Crete, P.O.
        Box 2208, GR-710 03 Heraklion, Crete, Greece} }
       
\begin{document}

\begin{abstract}
In two space-time dimensions, there is a theory of Lorentzian 
quantum gravity which
can be defined by a rigorous, non-perturbative path integral and is
inequivalent to the well-known theory of (Euclidean) quantum Liouville
gravity. It has a number of appealing features:  
i) its quantum geometry is non-fractal, ii) it remains consistent
when coupled to matter, even beyond the c=1 barrier, iii) it is 
closer to canonical quantization approaches than previous path-integral
formulations, and iv) its construction generalizes to higher dimensions.
\end{abstract}

\maketitle

\section{Motivation}

The ultimate aim of the work described below is to learn more
about four-dimensional quantum gravity by relating 
non-perturbative canonical and covariant approaches, which
so far have not been successful separately.

By `covariant' we do not mean semi-classical gravitational
path integrals, but genuine ``sums over all metrics", which
usually involve a discretization of space-time. 
A prototype of this ansatz is quantum Regge calculus. 
With the help of numerical simulations, one tries to find
a non-trivial fixed point and an associated continuum theory 
of quantum gravity. A great deal of numerical expertise has been 
gathered in the approach of dynamical
triangulations, a recent variant of the Regge method.
Unfortunately, all investigations so far have concentrated on
path integrals for unphysical space-time metrics of {\em
Euclidean} signature. Unlike for some fixed background metrics,
there is no prescription of how to ``Wick-rotate" a general
Euclidean metric to Lorentzian signature. 

On the other hand, a lot of progress has been made in the last
ten years in an analytic formulation of {\em canonical} quantum gravity
based on a reformulation in terms of gauge-theoretic variables,
called ``loop quantum gravity". Although {\em a priori} based in
the continuum, the
quantum theory has a number of discrete features reminiscent
of a generally covariant version of a lattice gauge field theory.
However, in this approach some basic obstacles remain in 
defining a satisfactory quantum Hamiltonian evolution, and 
efficient numerical methods have not yet been developed.

It is tempting to try to combine the positive aspects of both
approaches, but one soon realizes that in order to relate the
two, a number of technical and conceptual 
difficulties have to be overcome. To narrow this gap, we want to 
define a {\em Lorentzian} path integral where individual
regularized space-time geometries in the sum are required to be causal,
reflected in a local ``light-cone structure" and the absence of
closed time-like curves. It should be appreciated that it is
relatively easy to write down Feynman sums of amplitudes
\begin{equation}
\sum_{{\rm causal \; geometries}\;\{ I\} } 
e^{i S^{\rm Einstein}(I)},\label{sum}
\end{equation}
but that it is very hard to construct concrete models with a
suitable regularization, such that the sum can be performed and
leads to a non-trivial continuum theory.

\section{An ideal testing ground: d=2}

The difficulties associated with defining the sum (\ref{sum}) can
be overcome, at least in dimension $d\equ 2$. There exists already a
rigorous discretized path integral for {\em Euclidean} geometries,
obtained by the method of dynamical triangulations, where the
path-integral sum is performed over all possible triangulations
$T$ (i.e. gluings of 
equilateral triangles). The 2d gravity action for fixed space-time
topology reduces to the cosmological-constant term
\begin{equation}
S=\lambda \int d^2 x\, \sqrt{|\det g |},
\label{action}
\end{equation}
for both Euclidean and Lorentzian metrics $g_{\mu\nu}$. After the
discretization, this term becomes proportional to $\lambda N(T)$,
with $N(T)$ counting the number of triangles contained in $T$.
The Euclidean state sum is given by
\begin{equation}
Z^{\rm eu}(\lambda )=\sum_N e^{-\lambda N} Z^{\rm eu}(N)=
\sum_N e^{-\lambda N}\sum_{T^{(N)}} 1.
\label{zeucl}
\end{equation}
With the help of ingenious combinatorial methods the 
counting of all triangulations $T^{(N)}$ of volume $N$ in the
sum on the right can be done
explicitly. Moreover, there is good evidence that the
method is diffeomorphism-invariant, since it reproduces the
results of continuum Liouville gravity in the continuum limit.
\begin{figure}[t]
\centerline{\epsfxsize=8.0cm\epsfysize=5.333cm\epsfbox{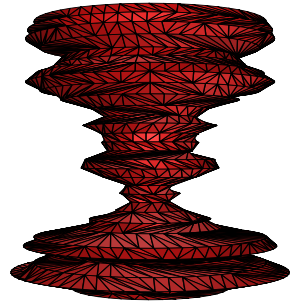}}
\caption{A typical triangulated 2d Lorentzian space-time 
(of topology $[0,1]\times S^{1}$).}
\end{figure}

How can this framework be adapted to the Lorentzian situation?
We have substituted the fundamental equilateral building blocks
(with squared edge lengths $a^2\equ 1$) by triangles with two
time-like edges with $a^2\equ -1$ and one space-like edge with 
$a^2\equ 1$ \cite{pap1}.
To obtain allowed histories, these must be glued causally: 
consecutive spatial slices (consisting entirely of space-like
edges) of variable length $l$ are connected by sets of time-like
edges. For simplicity, these slices are compactified to circles
$S^1$. A typical triangulated 2d Lorentzian geometry of $t$ 
time-steps ($t$ pointing up) is depicted in Fig.1. Note that the
local geometric degrees of freedom (apart from the edge lengths)
are encoded in the variable coordination numbers of edges meeting
at vertices, giving a direct measure of curvature.
It turns out that also in this discrete Lorentzian model, the 
combinatorics can be solved explicitly and yields the Lorentzian 
analogue $Z^{\rm lor}( \lambda )$ of (\ref{zeucl}). The partition
function exhibits critical behaviour as $\lambda\rightarrow
\lambda_{\rm crit}$, where a continuum limit can be taken.
After appropriate renormalization, one obtains a {\em new}
quantum gravity theory inequivalent to Liouville gravity.
It is rather surprising that there is a second universality class 
of models describing fluctuating two-geometries! 

The central dynamical quantity of the theory is the continuum 
propagator $G_\Lambda (L_1,L_2;T)$. It describes the
transition from an initial spatial geometry of length $L_1$
to a final one of length $L_2$ in proper time $T$ and takes
the form \cite{pap1}
\begin{eqnarray}
&&G_\Lambda (L_1,L_2;T)=e^{-\coth (\sqrt{i\Lambda} T) \sqrt{i
\Lambda} (L_1+L_2)}\nonumber\\
&&\hspace{1cm}
\times\frac{\sqrt{i\Lambda L_1 L_2}}{\sinh (\sqrt{i\Lambda} T)}\,
I_1\Big( \frac{2 \sqrt{i\Lambda L_1 L_2}}{\sinh (\sqrt{i\Lambda} T)}
\Big) ,\label{prop}
\end{eqnarray}
where $I_1$ denotes the modified Bessel function.

In order to illustrate our claim that the Lorentzian quantum
gravity theory differs from Liouville gravity, let us look at
the behaviour of a simple observable. A good example is the
so-called Hausdorff dimension $d_H$, which contains information
about the bulk properties of the quantum geometry in the
ground state of the theory. It is measured by looking at
the volume $V\!\sim\! r^{d_H}$ of geodesic balls (discs in dimension 2)
of radius $r$. Liouville gravity has a fractal Hausdorff
dimension $d_H\equ 4$. This may be surprising at first, but has to do
with the fact that the dominant contributions to the path integral
are highly branched geometries, with many ``baby universes".
By contrast, in the Lorentzian theory we have $d_H\equ 2$, which is the
``canonical" dimension expected from na\"\i ve semi-classical
considerations. The difference arises because there are no baby
universes in Lorentzian gravity. At a point where a baby universe
branches off, the Lorentzian metric structure must
inevitably go bad, thereby violating causality.
This also implies that in Lorentzian gravity the topology of the
spatial slices cannot change. Note that this is exactly the situation
described by canonical approaches to gravity. 

\section{Coupling matter to Lorentzian gravity}

The discussion of the previous section suggests that the
geometry of Lorentzian quantum gravity is ``better" behaved
than its Euclidean counterpart. This is also illustrated by
Fig.1 (taken from a Monte Carlo simulation
of pure Lorentzian gravity). In spite of strong fluctuations
$\langle \Delta l\rangle$ of the length of spatial slices, the
geometry is still effectively two-dimensional. 
The geometry of the Lorentzian model therefore lies somewhere
in between the wildly fluctuating and fractal quantum geometry
of the Liouville model and that of a fixed classical two-dimensional
space-time. 

It is an interesting question how matter will behave under coupling
to the Lorentzian model. To investigate this issue, we have
considered a model of Ising spins with nearest-neighbour interaction.
Coupling this to {\em Euclidean} dynamical triangulations yields an
exactly soluble model of Euclidean gravity plus matter. Its
matter behaviour is governed by the critical exponents
\begin{equation}
\alpha=-1,\quad \beta=0.5,\quad \gamma=2,
\label{expo}
\end{equation}
characterizing the singularity structure of the specific heat, 
the spontaneous magnetization, and the magnetic susceptibility as 
functions of the bare Ising coupling constant $\beta_I$. 
This should be contrasted with the Onsager values of these
exponents found on fixed, flat lattices, which are given by
\begin{equation}
\alpha=0,\quad \beta=0.125,\quad \gamma=1.75.
\label{onsager}
\end{equation}

The partition function for Lorentzian gravity coupled to Ising spins 
$\sigma_i=\pm 1$ is the sum
\begin{equation}
Z(\lambda,\beta_I )=\sum_N e^{-\lambda N} \sum_{T^{(N)}} 
Z_{T^{(N)}}(\beta_I ),
\label{matter}
\end{equation}
where the partition function $Z_T(\beta_I)$ of the Ising model
on the Lorentzian triangulation $T$ is 
\begin{equation}
Z_T(\beta_I )=\sum_{\{\sigma_i\} }
e^{\beta_I/2 \sum_{\left\langle ij\right\rangle }\sigma_i\sigma_j}.
\label{ising}
\end{equation}
We have investigated this model by means of a high-T (that is,
small inverse temperature $\beta_I$) expansion and by Monte Carlo
simulations \cite{pap2}. An exact solution has not yet been constructed.
Note that eq.\ (\ref{matter}) describes the Euclidean sector
of Lorentzian gravity plus matter, i.e. with real weights and
therefore Euclidean values for the coupling constants. This is
the form suitable for numerical simulations. 
What we have found is that both methods agree with good precision
in their estimates of the critical matter exponents, which turn
out to be the Onsager exponents. The Hausdorff dimension of the
geometry is unaltered, $d_H\equ 2$, and the typical Monte-Carlo-generated
geometries look qualitatively similar to the ones in pure gravity.
There {\em are} effects of the gravity-matter coupling at the discretized 
level, for example, on
the distribution of coordination numbers, but we have not investigated
whether this is reflected in a change of universal properties of
the geometry that would survive in the continuum limit. 

\section{Coupling more matter}

The previous picture is changed drastically when several Ising
models instead of one are coupled to Lorentzian gravity. For
the case of $n$ Ising models, the partition function (\ref{matter})
is replaced by 
\begin{equation}
Z(\lambda,\beta_I )=\sum_N e^{-\lambda N} \sum_{T^{(N)}} 
Z^n_{T^{(N)}}(\beta_I ).
\label{morematter}
\end{equation}
At the critical point, this model describes a conformal field
theory with central charge $c\equ n/2$ coupled to gravity.
Our motivation for coupling more matter is the fact that
{\em Euclidean} 2d gravity becomes inconsistent for $n>2$, that is,
beyond the so-called $c\equ 1$ barrier. In the presence of Ising spins
it is energetically favourable to have short boundaries between
regions of opposite spins. In a theory of fluctuating geometry
the effect of the spins is to try and ``squeeze off" parts of
the space-time manifold. In Euclidean gravity, where the geometry
is very branched to start with, this mechanism seems to be so
effective that for $n>2$ the theory seizes to make sense.

In order to get a clear picture of what goes on ``well beyond
the $c\equ 1$ barrier" in Lorentzian gravity, we have investigated
its properties at $n\equ 8$ by numerical simulations \cite{pap3}. 
One observes a very strong interaction of gravity and matter, to the
extent that the geometry is now in a different phase from before:
time and space directions acquire an anomalous relative
scaling and the Hausdorff dimension is changed to $d_H\equ 3$!
This is illustrated by Fig.2. The effect of the matter on the
geometry is reflected in the presence of the long, stalk-like
part of the space-time, which is effectively one-dimensional.
All interesting physics (that survives the limit as $N\!\rightarrow
\!\infty$) happens in the extended bulk phase. However, in spite
of these drastic changes in the geometrical properties, we have
found that the critical matter exponents retain their Onsager values!

\section{Conclusions}

There are a number of lessons to be learned from this 
two-dimensional model of quantum gravity. The choice of
Lorentzian over Euclidean, which in our case consisted
in the imposition of a causality condition
on individual path-integral histories, made a big
difference. In two dimensions, it led us to the discovery
of a new universality class of quantum gravity models,
besides that of Liouville gravity. 
In Lorentzian gravity, the quantum geometry is much smoother,
and better behaved in the sense that one can cross the infamous
$c\equ 1$ barrier without any problems. Conversely, the coupled
model with eight Ising models illustrated that the
matter behaviour is rather robust: the geometry can undergo
drastic changes without the critical matter behaviour
being affected. From this we also learn that Onsager
exponents by no means imply that the underlying space-time
is flat. 

The difference between the Euclidean and Lorentzian theories
can be traced entirely to the presence of branchings or
baby universes \cite{pap1,pap3}. Since this is a purely
kinematical effect which has to do with an {\em a priori}
restriction on the sum-over-geometries, it will be
present in higher dimensions as well. To date, the problem with
dynamically triangulated path integrals for
Euclidean geometries in $d\! >\! 2$ has been the dominance of
highly degenerate geometries, including a
proliferation of baby universes. Our hope is that also
in these cases a causality requirement will lead to an effective
``smoothing out" of the quantum geometry. An investigation
of the case $d\equ 3$ is under way.

\begin{figure}[t]
\centerline{
\epsfxsize=8.0cm\epsfysize=5.333cm\epsfbox{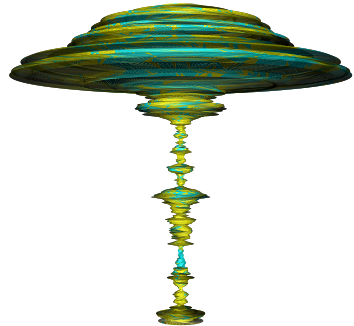}}
\caption{Lorentzian gravity coupled to 8 Ising models (N=18816,
t=168).}
\end{figure}


\begin{thebibliography}{9}
\bibitem{pap1} J. Ambj\o rn, R. Loll, Nucl. Phys. B536 (1998) 407-434
[hep-th/9805108].
\bibitem{pap2} J. Ambj\o rn, K. Anagnostopoulos, R. Loll, Phys. Rev. D,
to appear [hep-th/9904012].
\bibitem{pap3} J. Ambj\o rn, K. Anagnostopoulos, R. Loll, Phys. Rev. D,
to appear [hep-lat/9909129] and {\em preprint} AEI-1999-20 
[hep-lat/9908054].
\end{thebibliography}
\end{document}